\documentclass[twoside,12pt]{article}
\usepackage{epsfig}
\usepackage{amssymb,amsbsy,amsmath,amsfonts}
\usepackage{graphicx}
\usepackage{epsf,epsfig,float,latexsym,amsthm,fancyhdr,rotating}
\usepackage{graphics,psfrag,longtable}

\newcommand{\be}{\begin{equation}}
\newcommand{\ee}{\end{equation}}
\newcommand{\bea}{\begin{eqnarray}}
\newcommand{\eea}{\end{eqnarray}}

\begin{document}
\title{Relativistic chiral representation of
 the $\pi N$ scattering amplitude II:\\
 The pion-nucleon sigma term}
\author{J. Martin Camalich$^1$, J.M. Alarc\'on$^{2}$,
J.A. Oller$^{2}$,\\
$^1$Department of Physics and Astronomy, University of Sussex, BN1 9QH, Brighton, UK\\
$^2$Departamento de F\'isica. Universidad de Murcia. E-30071, Murcia, Spain}
\maketitle
\begin{abstract} 
We present a determination of the pion-nucleon sigma-term based on a novel analysis of the $\pi N$ scattering amplitude in Lorentz covariant baryon chiral perturbation theory renormalized in the extended-on-mass-shell scheme. This amplitude, valid up-to next-to-next-leading order in the chiral expansion, systematically includes the effects of the $\Delta(1232)$, giving a reliable description of the phase shifts of different partial wave analyses up to energies just below the resonance region. We obtain predictions on some observables that are within experimental bounds and phenomenological expectations. In particular, we use the center-of-mass energy dependence of the amplitude adjusted with the data above threshold to extract accurately the value of $\sigma_{\pi N}$. Our study indicates that the inclusion of modern meson-factory and pionic-atom data favors relatively large values of the sigma term. We report the value $\sigma_{\pi N}=59(7)$~MeV.    
\end{abstract}

The sigma terms, $\sigma_{\pi N}$ and  $\sigma_s$, are observables of fundamental importance that embody the internal scalar structure of the nucleon, becoming an essential piece to understand the origin of the mass of the ordinary matter in the context of non-perturbative QCD. The pion-nucleon sigma term, $\sigma_{\pi N}$, also plays a role in the study of the equation of state of the nuclear and neutronic systems \cite{NM:EOS} and is a key ingredient in investigations of the QCD phase diagram that explore the restoration of chiral symmetry in cold and dense matter \cite{NM:ChRest}. On the other hand, $\sigma_{\pi N}$ and  $\sigma_s$, as properties quantifying the response of the nucleons as probed by scalar interactions, appear in the hadronic matrix elements of the spin-independent neutralino-nucleon elastic scattering cross section~\cite{Bottinoetal,Ellis:2008hf,Giedt:2009mr}. Unfortunately, our current knowledge of these quantities is far from satisfactory and they have become the main source of uncertainty in the interpretation of experimental searches of supersymmetric dark matter~\cite{Bottinoetal,Ellis:2008hf}. With the advent of experimental results on direct searches of dark matter, Ellis \textit{et al} recently pled for a more accurate experimental determination of the sigma terms~\cite{Ellis:2008hf}. This has triggered an intense campaign for the obtention of these matrix elements from first principles using LQCD simulations~\cite{Young:2009ps}, although a revision of the experimental discrepancies is still missing. In this letter, we focus on the extraction of the $\sigma_{\pi N}$ from $\pi N$ scattering data. 

A connection between the $\pi N$ scattering amplitudes and $\sigma_{\pi N}$ can be established, using the chiral symmetry of the strong interactions, at the Cheng-Dashen point~\cite{Hohler:1982ja,Cheng:1970mx}, which lies in the unphysical region of the amplitude ($W_{cd}=\sqrt{s_{cd}}=M_N$, $t_{cd}=2 m_\pi^2$ with the physical region above $W_{th}=m_\pi+M_N$ and $t<0$). At this point, one has $\Sigma=\sigma_{\pi N}+\Delta_\sigma+\Delta_R$, where $\Sigma$ is proportional to the Born-subtracted isoscalar $\pi N$ scattering amplitude and the pion semileptonic decay constant squared $f_\pi^2$, $\Delta_\sigma\simeq15$~MeV~\cite{Gasser:1990ce} and $\Delta_R$ is the remainder of the relation which chiral symmetry constrains to be $\sim\mathcal{O}(m_\pi^4)$ and negligible ($\Delta_R\lesssim1$~MeV)~\cite{Gasser:1987rb}. The conventional way to perform the extrapolation from the physical to the subthreshold region is by means of an energy-dependent parameterization of the experimental data in partial waves (PW) supplemented by dispersion relations that impose strong analyticity and unitarity constraints from the high-energy data onto the low-energy scattering amplitude ~\cite{Hohler:1982ja}. There are two of such analyses to be considered in this letter: those of the Karlsruhe-Helsinki (KH)~\cite{ka85} and the George-Washington~\cite{wi08} (GW) groups. In particular, KH reports a $\Sigma=64(8)$~MeV that leads to $\sigma_{\pi N}=49(8)$~MeV~\cite{ka85}, while the analysis of Gasser \textit{et al}, based on the KH amplitudes, gives the classical result $\sigma_{\pi N}\simeq 45$~MeV~\cite{Gasser:1990ce}. On the other hand, the analyses of the GW group, including modern pion factory data, obtain larger values, $\Sigma=79(7)$~MeV that leads to $\sigma_{\pi N}=64(7)$~MeV~\cite{Pavan:2001wz}. 

The $\pi N$ scattering observables can be studied using chiral perturbation theory ($\chi$PT), which is the effective field theory of QCD at low energies~\cite{ChPT,Scherer:2002tk}. The advantage of $\chi$PT over the dispersive methods discussed above stems from the fact that this approach imposes the chiral Ward identities of QCD valid up to a certain order in the chiral power counting. Besides providing the corrections to the relation at the Cheng-Dashen point, it establishes a more direct connection between the isoscalar $\pi N$ scattering amplitude and $\sigma_{\pi N}$. In fact, up to $\mathcal{O}(p^3)$ both observables are connected by the same low-energy constant (LEC) $c_1$~\cite{Fettes,BecherLeutwyler}. The LECs are parameters that accompany the effective operators in the chiral Lagrangian and are fixed by matching to QCD or by comparison with experimental data. Therefore, in $\chi$PT one can predict $\sigma_{\pi N}$ fixing the relevant LEC using $\pi N$ scattering data and avoiding the extrapolation to the Cheng-Dashen point.

\begin{table}
\centering
\caption{Main physical observables obtained from the N$^2$LO $\pi N$ scattering amplitude in the EOMS renormalization scheme fitted to different PW analyses up to $W_{max}=1.2$~GeV ($W_{max}=1.16$~GeV for EM). Results of the scattering lengths are in units of 10$^{-2}$ $m_\pi^{-1}$.\label{Table:Observables}}
\begin{tabular}{ccccccc}
&$\chi^2_{\rm d.o.f.}$&$h_A$&$g_{\pi N}$&$\Delta_{GT}$ [\%] &$a_{0+}^+$ &$a_{0+}^-$ \\
\hline
KH~\cite{ka85}&0.75&3.02(4)&13.51(10)&4.9(8)&$-1.2(8)$&8.7(2)\\
GW~\cite{wi08}&0.23&2.87(4)&13.15(10)&2.1(8)&$-0.4(7)$&8.2(2)\\
EM~\cite{Matsinos:2006sw}&0.11&2.99(2)&13.12(5)&1.9(4)&0.2(3)&7.7(1)
\end{tabular}
\end{table}

In this note, we report on the extraction of $\sigma_{\pi N}$ from $\pi N$ scattering data and using a chiral representation scattering amplitude that is obtained in Lorentz covariant B$\chi$PT renormalized in the extended-on-mass-shell (EOMS) scheme. A detailed explanation of the advantages of this approach as compared with previous works has been presented by J.M.~Alarcon~\cite{JMTalk} in his talk at this School. We have performed a calculation up to N$^2$LO accuracy explicitly including the effects of the $\Delta(1232)$ resonance in the $\delta$-counting, that at low-energies and below the resonance region, takes into account the fact that the diagrams with the $\Delta$ are suppressed in comparison with those with the nucleon~\cite{Pascalutsa}. A more thorough presentation of our results can be found in~\cite{Alarcon:2011zs}. In the following, we outline some results of the resulting chiral representation on main observables and more specificaly on the pion-nucleon sigma term. 

The calculation in the EOMS scheme~\cite{EOMS} proceeds as explained in~\cite{JMTalk}. On top of that, we include Born-terms with an intermediate $\Delta(1232)$, whereas the corresponding loops with $\Delta$ propagators are of higher-order. We then fix the values of the LECs fitting the CM energy dependence of the 2 $S$- and 4 $P$-wave phase shifts obtained from the chiral amplitude to the latest solutions of the KH~\cite{ka85} and GW~\cite{wi08} groups. In addition, we include the analysis of the Matsinos' group (EM) ~\cite{Matsinos:2006sw} which focuses in the PW parameterization of the data at very low energies without imposing dispersive constraints from the high-energy region. The fits are done from the lowest CM energies above threshold, $W_{th}\simeq 1.08$~GeV, up to $W_{max}=1.2$ GeV which is below the $\Delta(1232)$ region (the EM analysis only reaches $W_{max}\simeq1.16$~GeV). The parameters $g_A$, $M_N$, $M_\Delta$, $m_\pi$ and $f_\pi$ are fixed to their experimental values $g_A=1.267$, $M_N=939$~MeV, $M_\Delta$=1232~MeV, $m_\pi=139$~MeV and $f_\pi=92.4$~MeV. We work with physical values which is equivalent to a reordering of the chiral series~\cite{IROurs}. The $N\Delta$ axial coupling can be determined using the $\Delta(1232)$ width, $\Gamma_\Delta=118(2)$~MeV, giving $h_A=2.90(2)$~\cite{Pascalutsa}. However, we also fit $h_A$ to compare the result with this value. 

\begin{figure}[t]
\centering
\psfrag{ss}{{\tiny $W$ [GeV]}}
\psfrag{S11per}{{\tiny$S_{11}$}}
\psfrag{S31per}{{\tiny $S_{31}$}}
\psfrag{P11per}{{\tiny $P_{11}$}}
\psfrag{P13per}{{\tiny $P_{13}$}}
\psfrag{P31per}{{\tiny $P_{31}$}}
\psfrag{P33per}{{\tiny $P_{33}$}} 
\epsfig{file=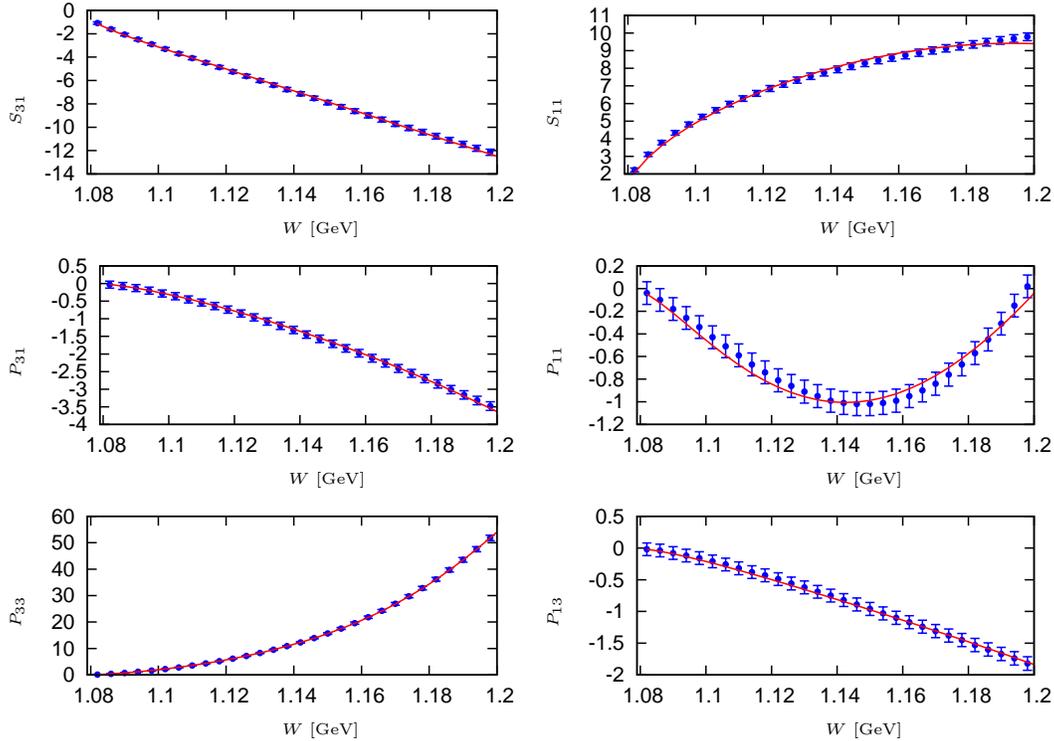,width=10cm,angle=-90} 
\caption{(Color on-line) Phase shifts given by the Lorentz covariant N$^2$LO $\pi N$ scattering amplitude in the EOMS  scheme fitted to the GW solution (circles)~\cite{wi08} up to $W_{max}=1.2$~GeV. \label{fig_graph}}
\end{figure}

The main results on physical observables are shown in Table~\ref{Table:Observables}. The errors quoted there are only of statistical origin and additional theoretical uncertainties are to be added. We plot in Fig.~\ref{fig_graph} the phase shifts of the $S$- and $P$-waves given by the N$^2$LO $\pi N$ scattering amplitude in the EOMS scheme fitted to the GW solution (circles)~\cite{wi08} up to $W_{max}=1.2$~GeV. Similar plots can be obtained for the KH and EM solutions.  The figure shows that the description for the lowest PWs is very good up to energies below the $\Delta(1232)$ resonance. This is reflected by the small $\chi^2_{\rm d.o.f.}$ listed in the second column of Table ~\ref{Table:Observables}, showing that the description of GW and EM PW analyses is better. It is also worth remarking that the problems encountered by the IR prescription to describe the $P_{11}$ wave of the GW group ~\cite{IROurs} disappear in the present analysis.

As one can infer from the third column of Table~\ref{Table:Observables}, only the GW solution gives a result on $h_A$ that is perfectly compatible with the determination from the $\Delta(1232)$ width. In the fourth column, we show the values obtained for the $\pi N$ coupling that, compared with the axial coupling $g_A$, gives the GT discrepancy $\Delta_{GT}$ in the fifth column (see Ref.~\cite{IROurs} for details). One can see that the results extracted from the KH, GW and EM PWs are completely consistent with the values reported by these collaborations, $g_{\pi N}=13.41(14)$~\cite{ka85}, $g_{\pi N}=13.15(1)$ ~\cite{wi08} and $g_{\pi N}=12.98(12)$~\cite{Matsinos:2006sw} respectively. However, the KH result is in clear disagreement with the numbers independently extracted from $NN$-scattering ($g_{\pi N}\simeq13.0$)~\cite{deSwart:1997ep} and pionic atom ($g_{\pi N}=13.12(9)$)~\cite{Baru:2010xn} data, leading to a $\Delta_{GT}$ that may be considered too large~\cite{BecherLeutwyler}. Finally, our full results, with the explicit inclusion of the $\Delta$, confirm the findings shown in~\cite{JMTalk}. Namely, that the large violation of the GT discrepancy obtained in IR ($\sim$20\%) is a spurious effect introduced by this scheme rather than a problem of chiral convergence in the $\pi N$ system.

\begin{table}
\centering
\caption{Values of the $\mathcal{O}$($p^2$) LECs in units of GeV$^{-1}$ and of $\sigma_{\pi N}$ in MeV obtained from the different $\pi N$ PW analyses.
 \label{Table:ResSigma}}
\begin{tabular}{cccccc}
&$c_1$ &$c_2$ &$c_3$ &$c_4$&$\sigma_{\pi N}$\\
\hline
KH&$-$0.80(6)&1.12(13)&$-$2.96(15)&2.00(7)&43(5)\\
GW&$-$1.00(4)&1.01(4)&$-$3.04(2)&2.02(1)&59(4)\\
EM&$-$1.00(1)&0.58(3)&$-$2.51(4)&1.77(2)&59(2)
\end{tabular}
\end{table}

Results for the isoscalar ($a_{0+}^+$) and isovector ($a_{0+}^-$) scattering lengths are shown in the last two columns of Table~\ref{Table:Observables}. These scattering parameters are interesting because they can be compared with the values independently extracted from pionic-atom data. Moreover, $a_{0+}^+$, as a parameter enclosing information on the isoscalar scattering amplitude, is tightly connected with $\sigma_{\pi N}$~\cite{Gasser:1990ce}. In this sense, the latest analyses of pionic-atom data clearly favor positive values for this observable, $a_{0+}^+=0.0076(31)$ $m_\pi^{-1}$~\cite{Baru:2010xn}. As we can see by direct comparison with the values shown in Table~\ref{Table:Observables}, only the GW and EM analyses are compatible with  $a_{0+}^+>0$. The impact that the pionic-atom result for $a_{0+}^+$ has on the value of $\sigma_{\pi N}$ is addressed below. In case of the isovector length, KH and GW are consistent with the pionic-atom value $a_{0+}^-=0.0861(9)$~\cite{Baru:2010xn}. For a thoughtful discussion on the discrepancy of the isovector scattering length extracted from the low-energy scattering data by the EM analyses, $a_{0+}^-=0.0797(11)$ $m_\pi^{-1}$, see Refs.~\cite{Matsinos:2006sw}. 

We have explored the subthreshold region and obtained the subthreshold coefficients which result from a Taylor expansion of the amplitudes about the point $W_{cd}$ and $t=0$~\cite{Hohler:1982ja}. In general, we observe a good agreement on the leading coefficients as compared with those reported by the KH and GW solutions, what positively speaks of the potential of our amplitude to connect, in the CM energy $W$, the physical region to the Cheng-Dashen point. However, it is known that, at N$^2$LO, the chiral representation underestimates the extrapolation to $t=2 m_\pi^2$ which is dominated by the threshold to two pions~\cite{Gasser:1990ce,BecherLeutwyler}. Nonetheless, and as was mentioned above,  B$\chi$PT can predict $\sigma_{\pi N}$ exclusively from the dependence in the CM energy $W$ via the LEC $c_1$ fitted to scattering data and avoiding the extrapolation into the unphysical region.
  
The expression for $\sigma_{\pi N}$ can be obtained either from the scalar form factor or the quark-mass dependence of the nucleon mass. The result in EOMS-B$\chi$PT is
\begin{eqnarray}
\sigma_{\pi N}=-4c_1 m_\pi^2-\frac{3g_A^2 m_\pi^3}{16\pi^2f_\pi^2 M_N}\left(\frac{3M_N^2-m_\pi^2}{\sqrt{4M_N^2-m_\pi^2}}\arccos{\frac{m_\pi}{2M_N}}
+m_\pi\log{\frac{m_\pi}{M_N}}\right).\label{Eq:SpiN}
\end{eqnarray}          
Eq.~(\ref{Eq:SpiN}) is valid up to NLO accuracy in the chiral expansion and has an uncertainty from higher-order contributions, which we  estimate computing the next subleading correction coming, in the $\delta$-counting, from the loop with an insertion of the $\Delta$~\cite{Pascalutsa}. This amounts to a contribution of $\sim-6$~MeV to be compared with the NLO term of $-19$~MeV. It is important to stress that we take this correction as an irreducible uncertainty in our determination. In order to explicitly add this contribution one has to include the same type of terms, arising from loops with one insertion of a $\Delta$ propagator, into the $\pi N$ scattering amplitude studying the changes produced in the LECs. 

To obtain the values of $\sigma_{\pi N}$, we consider fits to the KH and GW PW phase shifts and for various $W_{max}$, from $1.14$~GeV to $1.2$~GeV in intervals of $0.01$~GeV. For the analysis of EM we perform the same study up to its maximum CM energy $\sim1.16$~GeV. The purpose of this strategy is to take into account the dispersion of $\sigma_{\pi N}$ against the data set included in the fits. In Table~\ref{Table:ResSigma}, we show the mean and mean deviation for $c_1$ and $\sigma_{\pi N}$. We also include results for the other $\mathcal{O}$($p^2$) LECs for comparison purposes. In general, we find that for the KH and GW analyses $\sigma_{\pi N}$ decreases from 48~MeV to 39~MeV and from 65~MeV to~54 MeV respectively, when we increase $W_{max}$ within the given bounds. The results for the EM solution remain quite stable. As we can see, the values of $\sigma_{\pi N}$ are not completely consistent among each other. The number obtained from the KH PWs agrees with the canonical result $\sigma_{\pi N}~\simeq45$~MeV, whereas those of the GW and EM solutions completely agree with each other, and with the one determined by the GW group, in a value $\simeq16$~MeV larger. 

The consistency between the results derived from the GW and EM solutions is very remarkable since these are quite different analyses having both in common the implementation of the wealth of low-energy data collected along the last 20 years in meson factories~\cite{wi08,Matsinos:2006sw} with many points not included in KH~\cite{ka85}. Besides that, the positive values determined from pionic-atom data for $a^+_{0+}$, which are consistent with those extracted from the GW and EM PWs, strongly constrain $\sigma_{\pi N}$, raising it for $\gtrsim$7~MeV as compared with the KH values~\cite{Pavan:2001wz,Gasser:1990ce}. We also stress that only our results based in the GW PW analysis are perfectly compatible with all the phenomenology discussed. With these considerations, one may obtain the following value for $\sigma_{\pi N}$, as it is extracted from the analysis of $\pi N$ modern scattering data and using B$\chi$PT,
\begin{equation}
\sigma_{\pi N}=59(7) {\rm MeV}.
\end{equation}
The error includes the higher-order uncertainties estimated above added in quadrature with those given by the dispersion of the values in the average of the GW and EM results. If one were to include the KH result in this estimation, the result would by slightly reduced by 2-3~MeV.

As a concluding observation we want to address the fact that this large result on $\sigma_{\pi N}$ might be in conflict with some established phenomenology. In particular, it resurrects an old puzzle concerning the strangeness content of the nucleon~\cite{Pavan:2001wz}. This relies on the relation that is obtained in HB$\chi$PT up to NLO accuracy among the $SU(3)_F$-breaking of the baryon-octet masses, $\sigma_{\pi N}$ and the observable $y$ quantifying the strangeness content of the nucleon~\cite{Gasser:1982ap}. For the value of the sigma term obtained in the present work, this relation leads to a contribution of the strange quark to the nucleon mass of several hundreds of MeV. This is a physical scenario that is very hard to understand. In this regard, it has been recently shown that the HB framework does not provide a reliable description of $SU(3)_F$-breaking phenomenology and, in particular, of the quark-mass dependence of the baryon octet masses~\cite{MartinCamalich:2010fp}. In fact, in the analysis of LQCD data done using Lorentz covariant B$\chi$PT (EOMS scheme) and including the decuplet-baryon contributions, it has been obtained that a $\sigma_{\pi N}$ of $\sim$60~MeV is perfectly compatible with a negligible strangeness in the nucleon~\cite{MartinCamalich:2010fp}. Therefore, and in our opinion, the relation between the $SU(3)_F$-splitting of the baryon octet masses and the sigma terms ought to be revised.

Another caveat for a relatively large value of the pion-nucleon sigma term arises in chiral approaches of nuclear matter~\cite{NM:EOS,NM:ChRest}, in which $\rho\sigma_{\pi N}$ controls the leading contribution in the density ($\rho$) dependence of the quark condensate. At this order, a large value for $\sigma_{\pi N}\simeq 60~$MeV would imply a vanishing in-medium quark condensate  at $\simeq2\rho_0$ (with $\rho_0$ the nuclear saturation density), while this occurs at $\simeq3\rho_0$ using the canonical value of 45~MeV~\cite{NM:EOS}. However, a non-vanishing value of the in-medium temporal component of the pion axial coupling, $f_t$, is also a necessary condition for the spontaneous chiral symmetry breaking~\cite{OllerMedium}. The leading in-medium contribution to $f_t$ is controlled by the combination $c_2+c_3$, which is not directly related to $\sigma_{\pi N}$ at this order. The difference in $c_2+c_3$ between KH and GW sets in Table~\ref{Table:ResSigma} is only of 10$\%$, which means that the running with density for $f_t$, as obtained in~\cite{OllerMedium} (which also includes $NN$ interactions), differs also on a $10\%$. This figure is much smaller than the difference of around a $30\%$ in the density dependence of the quark condensate due to the different values of $\sigma_{\pi N}$~\cite{OllerMedium}, so that the vanishing of $f_t$ occurs at $\simeq3.5\rho_0$ for both PW analyses. A more thorough analysis is then necessary to put in agreement the simultaneous vanishing of $f_t$ and the quark condensate with density in order to properly discuss about chiral symmetry restoration in nuclear matter.

In summary, we have presented the values on $\sigma_{\pi N}$ resulting from a novel analysis of the $\pi N$ scattering amplitude in Lorentz covariant B$\chi$PT within the EOMS scheme up to ${\cal O}(p^3)$~\cite{JMTalk} and including the effects of the $\Delta(1232)$ explicitly in the $\delta$-counting. We have found that we perfectly describe the PW phase shifts of the KH, GW and EM groups up to energies below the $\Delta$-resonance region. Our amplitudes are suitable to extract an accurate value of $\sigma_{\pi N}$ from scattering data and avoiding the extrapolation to the unphysical region. Our work confirms the results obtained from dispersion relations by the KH and GW groups. This agreement, in turn, gives us further confidence in our analysis based on B$\chi$PT. We then ratify the discrepancy between them and give further support to the GW result after studying the latest EM solution. We conclude that modern data lead to a relatively high value of $\sigma_{\pi N}$.

This work is funded by the grants FPA2010-17806 and the Fundaci\'on S\'eneca 11871/PI/09. We also thank the financial support from  the BMBF grant 06BN411, the EU-Research Infrastructure Integrating Activity ``Study of Strongly Interacting Matter" (HadronPhysics2, grant n. 227431) under the Seventh Framework Program of EU and the Consolider-Ingenio 2010 Programme CPAN (CSD2007-00042). JMC acknowledges the MEC contract FIS2006-03438, the EU Integrated Infrastructure Initiative Hadron Physics Project contract RII3-CT-2004-506078 and the STFC [grant number ST/H004661/1] for support.

\end{document}